\documentclass[sigconf]{acmart}

\usepackage{graphicx,balance,multirow,xspace,subcaption,longtable}
\usepackage{tabularx,multirow,booktabs,blindtext}
\usepackage{makecell,amsmath,cleveref,rotating}
\usepackage[font={small}, textfont=md]{caption}
\usepackage[T1]{fontenc}
\usepackage[utf8]{inputenc}
\usepackage[draft,inline,nomargin,index]{fixme}
\usepackage{marginnote}
\usepackage{booktabs}
\usepackage{xcolor}
\usepackage{graphicx}
\usepackage{enumitem}
\usepackage{tcolorbox}

\tcbuselibrary{listingsutf8}

\setcopyright{none}

\AtBeginDocument{%
  }

\PassOptionsToPackage{hyphens}{url}
\usepackage{hyperref} 
\newcolumntype{P}[1]{>{\arraybackslash}p{#1}}
\newcolumntype{X}[1]{>{\centering\arraybackslash}p{#1}}

\expandafter\def\expandafter\UrlBreaks\expandafter{\UrlBreaks
  \do\a\do\b\do\c\do\d\do\e\do\f\do\g\do\h\do\i\do\j%
  \do\k\do\l\do\m\do\n\do\o\do\p\do\q\do\r\do\s\do\t%
  \do\u\do\v\do\w\do\x\do\y\do\z\do\A\do\B\do\C\do\D%
  \do\E\do\F\do\G\do\H\do\I\do\J\do\K\do\L\do\M\do\N%
  \do\O\do\P\do\Q\do\R\do\S\do\T\do\U\do\V\do\W\do\X%
  \do\Y\do\Z}

\crefformat{section}{\S#2#1#3}
\crefformat{subsection}{\S#2#1#3}
\crefformat{subsubsection}{\S#2#1#3}

\hypersetup{%
  pdftitle={YouTube Emotional Reinforcement},
  pdfauthor={},
  pdfkeywords={},
  bookmarksnumbered,
  pdfstartview={FitH},
  colorlinks,
  citecolor=black,
  filecolor=black,
  linkcolor=black,
  urlcolor=linkColor,
  breaklinks=true,
  hypertexnames=false
}

\newcommand\clearrow{\global\let\rowmac\relax}

\newcommand{\para}[1]{\vspace{.05in}\noindent\textbf{#1}}

\newcommand{\etal}{et al.\xspace}

\copyrightyear{2019} 
\acmYear{2018} 
\setcopyright{acmlicensed}
\acmConference[IMC '19]{acm}{imc}

\setcopyright{none}

\begin{document}

\renewcommand\footnotetextcopyrightpermission[1]{} 
\pagestyle{plain} 

\renewcommand{\sectionautorefname}{\S}
\renewcommand{\subsectionautorefname}{\S}
\renewcommand{\subsubsectionautorefname}{\S}


\title{YouTube Recommendations Reinforce Negative Emotions}
\subtitle{Auditing Algorithmic Bias with Emotionally-Agentic Sock Puppets}

\author{Hussam Habib}
\affiliation{%
  \institution{University of Iowa}
  \city{Iowa City}
  \country{USA}}
\email{hussam-habib@uiowa.edu}

\author{Rishab Nithyanand}
\affiliation{%
  \institution{University of Iowa}
  \city{Iowa City}
  \country{USA}}
\email{rishab-nithyanand@uiowa.edu}

\begin{abstract}
Personalized recommendation algorithms, like those on YouTube, significantly shape online content consumption.
These systems aim to maximize engagement by learning users' preferences and aligning content accordingly but may unintentionally reinforce impulsive and emotional biases.
Using a sock-puppet audit methodology, this study examines YouTube's capacity to recognize and reinforce emotional preferences.
Simulated user accounts with assigned emotional preferences navigate the platform, selecting videos that align with their assigned preferences and recording subsequent recommendations.
Our findings reveal reveal that YouTube amplifies negative emotions, such as anger and grievance, by increasing their prevalence and prominence in recommendations.
This reinforcement intensifies over time and persists across contexts.
Surprisingly, contextual recommendations often exceed personalized ones in reinforcing emotional alignment.
These findings suggest the algorithm amplifies user biases, contributing to emotional filter bubbles and raising concerns about user well-being and societal impacts.
The study emphasizes the need for balancing personalization with content diversity and user agency. 
\end{abstract}

\maketitle
\sloppy
\section{Introduction}
Recommendation algorithms shape a significant portion of the information we encounter online. On platforms like YouTube, recommended videos have become the primary source for content consumption, making content curation algorithms central to shaping our exposure to information. In recent years, YouTube's recommendation system has emerged as one of the most influential agents in our socio-political landscape. But it is also fueled by growing concerns about its role in disseminating controversial ideologies \cite{munger2022right, tufekci_opinion_2018}. One key feature that distinguishes modern platforms like YouTube from traditional media is their use of algorithms that dynamically adapt to users’ consumption behaviors and preferences. This is a crucial factor, as 76\% of videos watched on YouTube originate from recommendations. Similar to other content curation algorithms, YouTube's recommendation algorithm is designed to maximize user engagement—specifically, their watch time \cite{covington_deep_2016}. The algorithm achieves this by leveraging users' preferences, as revealed by their engagement behaviors. In practice, this means the algorithm serves content that aligns with users' revealed preferences; for example, a user who frequently watches basketball videos is more likely to receive additional basketball-related recommendations. However, researchers argue that optimizing for watch time and similar engagement metrics—which reflect users' immediate satisfaction—may not necessarily correlate with long-term satisfaction or the consumption of high-quality content \cite{kleinberg_challenge_2024}. Instead, such optimization might cause the algorithm to infer and reinforce impulsive preferences, which may not align with users' long-term well being \cite{cunningham_ranking_2023, cunningham_what_2024}.

It remains unclear how effectively recommendation algorithms optimized for watch time engagement metrics, such as YouTube’s, capture and reinforce users’ impulsive preferences, particularly their emotional biases. In this study, we investigate the capacity of YouTube's recommendation algorithm to recognize and reinforce users' emotional preferences.

\para{Emotions play a significant role in content engagement.} While content curation algorithms are complex and often opaque, the decision-making processes of individuals have proven even more difficult to understand—perhaps because they are often driven by subconscious biases. Although recommendation systems can expose users to undesirable content or create problematic feedback loops, individuals have independently demonstrated a tendency to disproportionately engage with such material, a phenomenon that predates and extends beyond the influence of these systems. For example, recent observational studies have uncovered problematic consumption behaviors driven by users \cite{chen_subscriptions_2023, hosseinmardi_causally_2024}. Since these recommendation systems learn from observed user behavior to predict future engagement, they inevitably learn and reinforce problematic consumption patterns. Emotional bias is a key driver of these consumption behaviors. Research indicates that content evoking high-arousal emotions is more likely to be shared and consumed extensively, making emotionally charged information inherently more engaging \cite{berger_what_2012, savolainen_delighting_2020, gupta_predicting_2019, de_leon_sadness_2021}. Furthermore, there are correlations between consuming emotionally intense content and negative outcomes such as radicalization, affective polarization, depression, and overall declines in mental well-being. While emotions play a significant role in content engagement, algorithms that indiscriminately amplify emotional biases risk promoting harmful consequences for individuals and society at large.

\para{YouTube might be optimizing for immediate satisfaction of the user.} In 2012, YouTube updated its recommendation algorithm to optimize for watch time rather than click-through rate. Initially, focusing on the click-through rate alone led to a rise in clickbait: misleading titles, suggestive thumbnails, and low-quality content. By shifting its focus from click-through rates to watch time, the platform significantly reduced these issues, resulting in a noticeable decrease in manipulative thumbnails and misleading titles. However, we continue along the same line of argument and contend that watch time is similarly a flawed proxy for user engagement. While optimizing for watch time addresses some issues associated with click-through rates, it does not necessarily correlate with content quality or long-term user well-being. Similar to how algorithms optimized for click-through rate reinforced clickbait and curiosity biases, we argue that optimizing for watch time reinforces emotional biases that are learned and reinforced by the algorithm.

\para{Our contributions.}
In this study, we investigate how YouTube's recommendation system responds to explicitly revealed emotional preferences. Specifically, we seek to determine whether YouTube has the capacity to capture and subsequently reinforce emotional biases. We conduct a sock-puppet audit of the recommendation algorithm, using automated bots with assigned emotional preferences—expressed through user engagement on the platform. By carefully controlling the revelation of these preferences, recording the utility of subsequent recommendations, and comparing outcomes with those of counterfactual control bots, we evaluate YouTube's tendency to reinforce emotional biases. In addition to this, we examine how personalized recommendations (informed by user watch history) compare with contextual recommendations (lacking user watch history and thus, revealed preferences) in reinforcing these biases. Finally, we explore whether the algorithm maintains the reinforcement across different recommendation contexts and whether this reinforcement intensifies over time.
Formally, in our work, we test the following hypotheses:

\begin{enumerate}[label=\textbf{H\arabic*.}, ref=H\arabic*]
    \item \textit{YouTube reinforces meaningful emotional preferences.} We hypothesize that revealing emotional preferences, by selecting preference-aligned content from within the recommendations, significantly increases the alignment of subsequent recommendations compared to the baseline of control bots. If valid, it would suggest the algorithm is able to recognize and reinforce emotional preferences. To test our main hypothesis, we operationalize it as the following two sub-hypotheses:
    
      \textbf{H1.1}: Up Next recommendations curated for preference-revealing sock puppets exhibit higher-than-baseline alignment with revealed preferences.

    \textbf{H1.2}: The correlation between alignment and rank of recommendations in the Up Next recommendations is stronger for preference-revealing sock puppets than for control sock puppets.

    \item \textit{YouTube reinforces emotional preferences through personalization.} We hypothesize that personalized recommendations reinforce users' emotional preferences more strongly than contextual recommendations do when users are viewing preference-aligned videos. Personalized recommendations are tailored to individual past behavior and revealed preferences, whereas contextual recommendations are based primarily on the content being currently viewed and lack user watch history information. If valid, it would suggest that the revealed emotional preferences are being learned to then be used for future personalization—beyond what would be expected from aggregate user behavior alone.

    \item \textit{YouTube amplifies exposure to emotionally aligned content.} We hypothesize that once the recommendation algorithm learns a user’s emotional preferences, it amplifies exposure to content aligned with those preferences. We define amplification as encompassing two phenomena: (1) the intensification of preference reinforcement over time and (2) the persistent reinforcement of revealed preferences. If valid, this would demonstrate that exposure is biased towards emotionally aligned content, driven by more than just immediate user behavior. To test this hypothesis, we propose the following sub-hypotheses:

    \textbf{H3.1}: Personalized Up Next recommendations become increasingly more reinforcing with continued engagement with preference-aligned videos, indicating an intensification of reinforcement over time.

    \textbf{H3.2}: Up Next recommendations, even for pre-selected videos not selected for their emotional alignment, continue to exhibit greater alignment with the user’s revealed preferences than the control group.
\end{enumerate}

\noindent Together, our hypotheses investigate the reinforcement of emotional preferences on YouTube. Through a carefully designed experiment, we demonstrate how personalization and aggregate user behavior shape the reinforcement, its amplification, and whether the reinforcement persists across unrelated videos. Finally, combining previous findings from audits of recommendation algorithms, we offer new insights into how recommendation algorithms shape exposure to information as users' emotional preferences are revealed and learned.

\section{Background} \label{sec:background}
In this section, we expand on our overview of the YouTube recommendation algorithm (\Cref{sec:background:algorithm}), emotional biases in user consumption behaviors (\Cref{sec:background:user-behavior}), and prior works (\Cref{sec:background:related-works}).

\subsection{YouTube's Recommendation Algorithm}
\label{sec:background:algorithm}
In this section, we outline YouTube's recommendation system as understood from prior literature and documentation.

\subsubsection{How do recommendation algorithms optimize for engagement?}
Fundamentally, optimizing for engagement involves learning users' preferences and providing recommendations that align with them. This approach is grounded in theories from behavioral economics, which suggest that individuals are more likely to engage with content that aligns with their preferences. Early recommendation systems attempted to capture these preferences through explicit user inputs, such as selecting topics or categories. However, these methods were limited in scope and accuracy because users often find it challenging to accurately articulate their own preferences. Modern recommendation algorithms overcome this limitation by inferring preferences from users' interaction histories—specifically, their choices and engagement patterns with content. This approach relies on the concept of revealed preferences, where the algorithm deduces what users prefer based on their behavior rather than their stated or real interests. The recommendation system then selects content that aligns with users' revealed preferences.

\para{Formally Representing the Optimization for Engagement on YouTube}
YouTube's recommendation problem can be formally modeled as predicting the probability that a user will watch a specific video, given their embedding and context. This model was described by YouTube engineers in their 2016 paper \cite{covington_deep_2016}. The probability can be represented by the following equation:
\[
P(w_t = i \mid U, C) = \frac{e^{v_i^T u}}{\sum_{j \in V} e^{v_j^T u}}
\]

Where \( w_t \) represents the video watched at time \( t \), \( U \) and \( C \) are the user and context embeddings. The user embedding $U$ is a high-dimensional representation of the user's revealed preferences, capturing their watch history, search behavior, and demographic information. The context embedding captures information such as the current video being watched, the time of day, the type of device, and other contextual features. Together, the user and context embeddings $u$, where \(u \in \mathbb{R}^N\), represent the revealed preferences and situational factors influencing the recommendation. The probability of watching video \( w_t = i \), given the user and context embeddings, is computed over all videos in the corpus of candidate videos \( V \). In this model, \( v_i \) represents the embedding vector for video \( i \) characterizing the content information, source information and engagement history of the video. \( u \) is the high-dimensional embedding of the user and context pair, where \( u \in \mathbb{R}^N \). The dot product between \( v_i \) and \( u \) measures the similarity between the candidate video and the combined user-context embedding. The model includes a softmax function to normalize these similarity values across all possible videos in the set, ensuring the output is a valid probability distribution over all candidate videos. This means that the likelihood of a particular video being watched depends on how well its embedding aligns with the user's preferences and the current context. Consequently, the video with the highest probability is the one most likely to be recommended for the user to watch next. In other words, the probability $P(w_t = i \mid U, C)$ reflects the model's estimation of how likely the user is to watch video \textit{i} next.

\para{Contextual recommendations.} The user embedding ($U$), which includes the watch history, influences the selection of the next recommended video, ensuring it aligns with the user's preferences. However, in the absence of the user embedding, such as when a user is not signed-in, the recommendation system relies on the context embeddings ($C$) and the video embeddings ($v_j$). We call recommendations generated in this scenario contextual recommendations. Notably, these recommendations are driven by collaborative and content-based filtering techniques based on the aggregate behavior data from prior users.

\subsection{Emotional preferences in consuming videos.}\label{sec:background:user-behavior}
Media that evokes strong emotions is more likely to become viral \cite{kitroeff_why_2014}, aligning with uses and gratification theory, which recognizes emotional needs as key drivers of media consumption. Researchers note that content appealing to such needs or evoking intense emotions increases viewer engagement, a phenomenon well studied in viral media research.

\subsubsection{Emotions can shape users' media consumption patterns.} \label{sec:background:user:selection}
How users choose which media content to engage with has long been a central question in communication studies. Traditional theories, such as uses and gratifications \cite{ruggiero_uses_2000}, suggest that people select media that align with their immediate psychological or social needs, including the desire for information, entertainment, and emotional release \cite{valentino_is_2008}. Increasingly, however, research indicates that the emotional tone of the content itself is a crucial factor driving its consumption \cite{robertson_negativity_2023} \cite{robertson_negativity_2023, fan_anger_2014, brady_emotion_2017, berger_arousal_2011, crockett_moral_2017, brady_mad_2020, carpenter_political_2020, ecker_psychological_2022, rathje_out-group_2021, baumeister_bad_2001}. Simply put, emotions evoked by a piece of media such as anger, sadness, or joy, significantly influence whether it's selected and engaged with. In this section, we explore emotions, sentiments, and cognitive constructs that are found to be drivers for media consumption. By examining prior work exploring their unique roles, we can better understand the mechanisms through which emotional content impacts user behavior.

\begin{itemize}
    \item \textit{Anger} amplifies online propagation and boosts virality; perceived anger in content can predict its eventual reach within social networks \cite{berger_what_2012, fan_anger_2014, brady_emotion_2017, berger_arousal_2011, crockett_moral_2017, brady_mad_2020, carpenter_political_2020, ecker_psychological_2022}.
    \item \textit{Grievance}, including feelings of sadness and discontent, prompts users to seek content that resonates with or helps regulate these emotions. Vulnerable individuals, in particular, gravitate toward content that expresses grievance, which may heighten susceptibility to harmful behaviors and ideologies \cite{habib_making_2022, noauthor_st_nodate, mccauley_mechanisms_2008, corner_mental_2018}.
    \item \textit{Group Identification}, a cognitive construct reflecting strong identification with specific groups, is another key factor. Content reinforcing in-group cohesion or highlighting out-group differences heightens polarization and deepens social divides. Individuals with strong group ties are more likely to engage with divisive materials \cite{rathje_out-group_2021, yardi_dynamic_2010}.
    \item \textit{Negative} sentiment captures attention, drives engagement, and fuels sharing, especially in political contexts, where it can sway public opinion and exacerbate polarization \cite{baumeister_bad_2001, brady_emotion_2017, robertson_negativity_2023, del_vicario_echo_2016}.
    \item \textit{Positive} sentiment, similar to strong negative emotions and sentient, also exerts substantial influence. By enhancing users’ moods and fostering social bonds, uplifting messages can spark increased sharing and engagement, illustrating that content need not rely solely on negative appeals.
\end{itemize}

\para{Influence on subsequent media consumption.} One of the significant and immediate outcomes of consumption of emotional content is its influence on subsequent selection of media. As highlighted in \cref{sec:background:user:selection}, one key factor in selecting and consuming media are emotions, specifically, the current emotional state of the individual. Within this altered state: consumers are more likely to continue engaging with content with similar emotional framing. For example, users consuming emotionally negative content are likely to continue consuming similar content, reinforcing the prevailing emotional state \cite{slater_reinforcing_2007}, constructing a self-reinforcing loop of emotional media consumption. This pattern demonstrates the self-reinforcing loop of emotional media consumption. 

\para{Influence on user behavior.}
Beyond immediate emotional impact, consuming emotionally charged content can have long-term effects that extend into broader contexts. For instance, exposure to anger-inducing content on social media may lead to heightened aggression in users' offline interactions \cite{goldenberg_amplification_2023}. Furthermore, strong negative emotions can shape how individuals perceive and understand their world; fear- and anger-inducing content may foster an "us versus them" mentality, increasing affective polarization \cite{martel_reliance_2020, brady_emotion_2017, knutson_news_2024}. For vulnerable individuals, the consequences o emotional influence can be particularly detrimental, potentially worsening their condition. This vulnerability is further exploited as emotional content frequently overlaps with problematic material, such as misinformation and biased information.
Studies demonstrate that false news often leverages emotional appeals to capture attention and drive engagement, with negative emotions like outrage and fear playing a particularly significant role \cite{martel_reliance_2020, brady_emotion_2017, knutson_news_2024} consequently spreading faster and wider. 

\subsection{Related works} \label{sec:background:related-works}
\subsubsection{Auditing YouTube’s recommendation algorithm.} A growing body of research has focused on auditing YouTube’s recommendation algorithm, particularly to understand its role in disseminating radical or extremist content. These investigations typically center on 1) whether the algorithm explicitly promotes extremist, radical, or fringe material, 2) whether users consume extremist recommendations, or 3) how strongly the platform pushes such content to its audiences. Constructing a causal analysis of how the algorithm contributes to online radicalization and related outcomes poses methodological challenges, therefore, different methodological approaches have yielded differing, yet related, conclusions. 

One line of methodology employs “sock-puppet” experiments, where controlled user accounts follow recommendation trails to evaluate how the algorithm steers viewers toward increasingly extreme content \cite{ribeiro_amplification_2023-1, haroon_youtube_2022, alfano_technologically_2021, brown_echo_2022}. The critical limitation of a sock puppet study is the inability to select videos from recommendations to watch as a user would, instead, experiment often traverse the recommendations tree on a predefined path (e.g., always select the first recommended video) or random walks. Notably, studies employing sock puppet studies on blank user accounts consistently show an amplification or filter bubble effect. For instance, Ribeiro et al.\cite{ribeiro_auditing_2020} document an increase in far-right content. Haroon \etal \cite{haroon_youtube_2022} and Alfano \etal \cite{alfano_technologically_2021} find similar outcomes in their experiments where recommendation nudges gradually guide users from mainstream topics to conspiracy theories or other forms of radical content. In addition to the random or predefined selection to traverse the recommendation tree, these studies often do not incorporate representative user history. In their work, Brown \etal found a limited filter bubble or rabbitholing effect when recruited users were instructed to follow a predefined path in the recommendations tree on their personal YouTube account. Recommendation systems are dynamic and evolve based on the content users select. Sock puppet studies with predefined pathways for user accounts to follow, regardless of blank or representative user history, limit their analysis to studying the recommendations rather than the recommendation system which requires meaningful selection of content.
On the other hand, observational studies based on user data suggest that the audience’s existing preferences play a substantial role in the filter bubble effect and the prevalent consumption of radical content \cite{chen_subscriptions_2023, hosseinmardi_causally_2024, hosseinmardi_examining_2021, munger2022right}. Specifically, studies analyzing the users online traces and browsing histories find users are more likely to engage with radical or extremist content than its prevalence in their recommendations \cite{chen_subscriptions_2023, hosseinmardi_examining_2021, munger2022right}. More recently, users' tendency to select more radical content than what is recommended to them has been confirmed by Hosseinmardi \etal through a combination of observational and sock puppet experiment. They found that users are more likely to choose radical content than the recommendations provided to them by YouTube’s algorithm

Overall, prior works auditing YouTube's recommendation algorithm study the responses of the system on arbitrary inputs (sock puppets studies) or uncontrolled inputs (observational studies) revealing how and when the algorithm serves problematic recommendations. In our work, we take control of inputs to the system, revealing assigned preferences dynamically, and record how these revealed preferences shape the subsequent recommendations. 

\subsubsection{Emotional content, virality, and emotional contagion.} Beyond radicalization, another line of research investigates how content curation systems—such as search engines and social media—amplify emotionally charged or moralized material. Because these platforms are frequently optimized for engagement, content that elicits strong reactions (e.g., anger, fear, or moral outrage) garners higher interaction and thus has a greater likelihood of recirculating among users. For instance, Crockett \etal \cite{crockett_moral_2017} illustrate how moral outrage in online environments can rapidly escalate collective emotional states, while Mcloughlin and North \cite{mcloughlin_misinformation_2024} connect heightened emotional reactions to the prevalence and spread of misinformation. Brady \etal~\cite{brady_emotion_2017} similarly find that emotion plays a pivotal role in the diffusion of moralized content, suggesting that engagement-optimized systems inherently favor emotionally charged information.

Such emotionally charged interactions can give rise to \emph{emotional contagion}, whereby emotions spread from user to user through digital platforms, heightening group-level emotional responses. This contagion effect is particularly salient on social media, where emotionally resonant messages can trigger rapid sharing, likes, and comments, effectively magnifying collective sentiment. Rathje \etal \cite{rathje_out-group_2021} highlight that content inducing out-group animosity yields spikes in social media activity, while Vosoughi \etal \cite{vosoughi_spread_2018} demonstrate that misinformation often travels faster than factual reporting—partially because of its emotional resonance. Recent work further explores how these virality-enhancing dynamics shape broader user attitudes and behaviors \cite{prollochs_emotions_2021, corbu_does_2020}, underscoring the interconnected roles of platform algorithms and user-driven engagement in amplifying both emotional content and the consequent contagion effects.
\section{Experiment Design}
\para{Overview.}
To uncover how YouTube's recommendation system responds to user preferences, we use a sock-puppet audit approach.
We create sock puppets ($U$)\textemdash simulated user accounts\textemdash that sign in to YouTube and maintain a persistent watch history. Sock puppet $U_{p}^{s}$ is assigned a preference $p \in \mathcal{P}$ and a seed video $s \in \mathcal{S}$ from a set of videos $\mathcal{S}$. 
To measure the alignment between videos and the assigned preference $p$, each sock puppet is equipped with a utility function $u_p(v)$,  quantifying how well a video $v$ aligns with the preference $p$.
Starting from the seed video $s$, the sock puppet reveals their assigned preference by selecting videos from the Up Next recommendationss list that most aligns with the assigned preference.
To formalize the sock puppet's behavior, we define a sequence of videos \(\{v_t\}_{t=0}^{100}\) where \(v_0 = s\) is the seed video. At each time step \(t \geq 0\), the sock puppet selects the next video \(v_{t+1}\) from the set of Up Next recommendations \(R(v_t)\) for the current video \(v_t\).  The selection is based on maximizing the utility function \(u_p(v)\) with respect to their assigned preference \(p\). 
This process ensures that the sock puppet consistently chooses videos that most closely align with their preference \(p\), thereby revealing their assigned preference through their viewing behavior.
Simultaneously, the sock puppets record all Up Next recommendations they receive during the process. In addition to this, they also periodically\textemdash after every $M=10$ steps\textemdash collect recommendations from a predefined set of videos $\mathcal{K}$.

In total, we repeat this procedure TTT times, using XX different preferences, YY seed videos, and ZZ repetitions for each preference–seed combination.

\subsection{Creating sock puppets.}
For each sock puppet, we create a Google Account that is signed into YouTube. These accounts are created through Google Workspace, a known practiced from prior works \cite{chandio_how_2024}. To allow sock puppets to instrument these accounts and interact with YouTube, we use YouTube's undocumented public API—InnerTube API \cite{luan_luanrtyoutubejs_2025}. Each sock puppet $U_{p}^{s}$ is assigned a preference $p \in \mathcal{P}$ and a seed video $s \in \mathcal{S}$ from a set of videos $\mathcal{S}$. From each seed category, 8 videos are randomly selected as seeds. Each preference-seed video is assigned two sock puppets to repeat and ensure statistical validity. In total, we create 560 sock puppets (7 preferences $\times$ 4 seed categories $\times$ 10 seed videos from each category $\times$ 2 replications)

\para{Using InnerTube API to interact with YouTube.} We use an undocumented YouTube API (known as the InnerTube API) that allows for unrestricted and unthrottled access to essential functionalities, such as adding videos to watch history, retrieving recommendations, and signing in. Other researchers have used this API to perform large-scale exploratory studies to analyze the platform's content libraries. In our work, we leverage the InnerTube API to conduct our sock puppet audits of the platform's recommendation algorithm. Specifically, we use three endpoints: adding a video to a user's watch history, getting recommendations for a video (which is independent of adding it to the watch history), and signing in. We discuss the InnerTube API and its consistency with desktop YouTube recommendations in the appendix \cref{sec:appendix:validation}.

\subsubsection{Assigning preference ($p$) to sock-puppets.}
Each sock puppet is assigned one of 7 preferences ($|\mathcal{P}| = 6 + 1$): six emptional and one control preference. Each preference has a corresponding utility function (\(u_p(v)\))  that measures the utility of a video, more specifically the transcript of the video, with respect to that preference $p$. The set of preferences used is informed by prior work and literature identifying prevalent emotional biases in online content, its propagation, and consumption \cref{sec:background:user-behavior}. A detailed description of each utility function is provided in Box 1. It is important to note that each function $u_p(v)$ operates on the transcript of video $v$ to produce a score reflecting the video's alignment with preference $p$.

\begin{tcolorbox}[float, colback=gray!5!white, colframe=gray!75!black, 
                  title=\textbf{Box 1:} Assigned Preferences and Utility Functions, 
                  label=box:preferences]
\small
Sock puppets are assigned one of seven preferences described below:
\vspace{0.2em}
\noindent
\begin{list}{\textbf{•}}{
    \setlength{\leftmargin}{0.5em}
    \setlength{\rightmargin}{0em}
    \setlength{\itemsep}{0.5em}
    \setlength{\parsep}{0em}
}
    \item \textbf{Anger}: Quantified using LIWC \cite{boyd_development_2022} to count words and phrases associated with anger in the video transcript. This measures the emotional intensity of anger expressed in the video.
    \item \textbf{Grievance}: Measured using LIWC \cite{boyd_development_2022} to analyze transcripts for words linked to sadness or grievance-related terms. This reflects the level of grievance conveyed in the video.
    \item \textbf{Group-Identification}: Detected through LIWC's \cite{boyd_development_2022} first-person plural pronouns, indicating in-group language, and third-person plural pronouns, indicating out-group language.
    \item \textbf{Negativity}: Computed with VADER \cite{hutto_vader_2014} sentiment analysis to quantify the level of negative sentiment in the video transcript, providing an average negativity score.
    \item \textbf{Positivity}: Similar to negativity computed using VADER sentiment \cite{hutto_vader_2014} analysis for positive sentiment, capturing the average positivity score of the video.
    \item \textbf{H-Frequency}: Selects videos based on an arbitrary textual feature (the proportion of the letter “h” in the transcript). This serves as a meaningless preference for comparison.
    \item \textbf{Random Position} (Control): Videos are selected randomly from the recommendation list without considering content alignment. This provides a baseline for analyzing the effects of preference-based selection.
\end{list}
\end{tcolorbox}

\subsubsection{Selecting seed videos ($s$)}
Each sock puppet begins with a single seed video; subsequent videos are selected from this seed video's recommendations. This seed video determines the universe of content the sock puppet will be exposed to. We create four distinct sets of seed videos, one random set of videos and three sets of videos from topical categories: \textit{Fitness}, \textit{Gaming}, and \textit{News} (see \cref{tab:videos} for sampled seed videos).

\para{Topical set of seed videos.}
To gain a clearer understanding of how reinforcement of preferences manifest within popular domains, we compile a set of seed videos from three widely consumed YouTube content categories: \textit{Fitness}, \textit{Gaming}, and \textit{News}. Each category is selected for its prevalence on the platform and its strong societal influence, ensuring that our chosen seeds are not niche sub-genres but rather familiar entry points into mainstream viewing habits. We collect 100 candidate seed videos by conducting keyword searches and collecting top results for each topic. From these candidate seed videos, we randomly select XX videos for each topic.

\para{Random set of seed videos.}
In addition to topic-based seeds, we also construct a random sample of videos. Following an approach developed \cite{zhou_counting_2011} and validated \cite{mcgrady_dialing_2023} by a prior work we generate a set of random, yet valid, YouTube video URLs. Unlike the topical approach, this method avoids any initial assumptions regarding a video’s theme or popularity. By repeatedly starting from content chosen entirely at random, we create a more unbiased baseline that better represents the platform’s content library. Over multiple iterations, this approach ensures that our results are not driven by specific genres, trending content, or popular channels, thus offering a more general perspective on how preferences influence recommendation patterns.

\subsection{Revealing assigned preferences.} \label{sec:data:reveal}
In this section, we describe how the sock puppets reveal their assigned preferences to YouTube and record both personalized and contextual recommendations during the experiment. We run the experiment in batches, where within each batch, all the sock puppets of different preferences are assigned an identical single seed video. Consequently, each batch has seven ($|\mathcal{P}|=6 + 1$) sock puppets, one for each preference and a control. The experiment begins by each sock puppet watching the seed video which automatically adds the video to YouTube's watch history. Next, the sock puppet requests the Up Next recommendations, a list of 20 videos. For each video $v \in R(v_t)$, the video transcript is requested and the utility $u_p(v)$ of the video is measured based on the sock puppet's preference. Due to limitations within our utility functions, our sock puppets (including control sock puppets) ignore videos that do not have a transcript or do not have an English transcript. In total, we observed a $YY\%$ rate of transcripts. The video $v^*_{t+1}$ with the highest utility for each sock puppet is selected and added to their watch history.
\[
v_{t+1} = \arg\max_{v \in R(v_t)} u_p(v).
\]
We refer to this video as \textit{preference-aligned} video.

\subsection{Recording recommendations}
The sock puppets record recommendations at two points: recommendations for preference-aligned videos—i.e., sock puppet selected videos—and recommendations and recommendations for a predefined set of videos. 

\subsubsection{Recording recommendations for preference-aligned videos.} \label{sec:data:record-aligned} While revealing its preference, sock puppets simultenously record all Up Next recommendations for the selected videos. This constructs the \{$R_{personalized(v_i)}$\}$^{N}_{i=1}$ dataset. These recommendations represent the immediate personalized response of the algorithm to the sock puppets' viewing behavior. In addition to the personalized recommendations, we also collect contextual recommendations ($R_{contextual}(v_i)$) for the same videos using a separate YouTube interface without signing in. These contextual recommendations provide a baseline for comparison, as they are influenced solely by the immediate context—the currently watched video—without personalization and watch history.

\subsubsection{Recording recommendations for predefined videos.}\label{sec:data:record-predefined} After every $M$ steps ($M = 20$), the sock puppets retrieve additional Up Next recommendations for a predefined set of videos. These predefined videos are identical for all sock puppets, regardless of their assigned preferences, starting seed, or viewing history. The recommendations from the predefined videos are represented as  $R_{personalized}(k_i)$ where $i$ represents the time steps when the recommendations are collected (every $M$ steps). This dataset allows us to isolate the influence of personalization without considering the context over time. The predefined videos include a mix of randomly obtained videos and videos related to the topics of \textit{fitness}, \textit{news}, and \textit{gaming}. This selection ensures coverage of a broad range of content, allowing us to observe the influence of personalization across different content domains.

\subsubsection{Summary of collected data.} \label{sec:data:summary}
For each preference-seed pair, we assigned two sock puppets to reveal their assigned preferences starting from the seed and capture subsequent Up Next recommendations. In total from each sock puppet, we collect 100 Up Next personalized recommendations for preference-aligned videos (selected by the sock puppet) in addition to their contextual recommendations. Additionally, we collect 5 sets of personalized recommendations for the pre-defined set of videos.
\section{Emotional reinforcement in personalized recommendations}
In this section, we test the hypothesis that YouTube's recommendation algorithm reinforces users' emotional preferences by curating recommendations that align with the user's revealed emotional preferences. Specifically, we compare the emotional utility of the Up Next recommendations shown to preference-revealing sock puppets against those shown their control counterparts. 
%
%
%
If valid, this hypothesis would suggest that the algorithm not only recognizes users' emotional biases but also sustains their exposure to emotionally aligned content.
Such reinforcement has serious implications for user experience, content production, and algorithms optimizing for engagement. 
Namely, we evaluate different metrics of reinforcement within the Up Next recommendations, i.e. prominence and prevalence of utility.  
\begin{enumerate}[label=\textbf{H1.\arabic*.}, ref=H1.\arabic*]
    \item \textit{Up Next recommendations curated for preference-revealing sock puppets exhibit higher alignment with revealed preferences compared to control recommendations.}
    This sub-hypothesis evaluates whether preference-revealing sock puppets receive recommendations with higher utility than control sock puppets, starting from the same seed.
    Validating this hypothesis would suggest that the algorithm actively sustains user biases, implying that the platform recognizes these emotional preferences.
    \item \textit{The correlation between utility and rank of recommendations in the Up Next recommendations is stronger for preference-revealing sock puppets compared to the control.} 
    This sub-hypothesis examines whether content with higher utility is ranked more prominently for sock puppets that reveal their preferences.
    If confirmed, this would demonstrate that YouTube's algorithm not only reinforces emotional preferences but also amplifies their prominence, potentially influencing user engagement patterns.
\end{enumerate}

\noindent Together, these sub-hypotheses help understand whether the algorithm reinforces emotional preferences as they are revealed, implying its capacity to recognize emotional preferences. 

\subsection{Methods}\label{sec:h1:methods}
To examine our hypotheses, we analyze personalized recommendations recorded by our sock puppets (see \cref{sec:data:record-aligned}) for preference-aligned videos, measuring their reinforcement and comparing reinforcement observed by preference-revealing sock puppets against their control counterparts.

\para{Estimating the baseline reinforcement.}
To establish emotional-reinforcement baseline within the domain of each seed video, we introduced control sock puppets.
The control sock puppets began with the same seed videos $s$ as their treatment counterpart, but made random selections rather than optimizing choices based on any assigned preference.
This baseline represents a counterfactual scenario, reflecting the natural reinforcement of recommendations within the seed domain without a revealed emotional bias.
%

\para{Operationalizing reinforcement of preferences as prominence and prevalence.} We used two key metrics to operationalize reinforcement:
\begin{itemize}
    \item  \textbf{Utility prevalence}: This measures the average utility score of all recommended videos exposed to a given sock puppet throughout the experiment, reflecting the overall alignment of recommendations with the assigned preference.
    \item  \textbf{Utility prominence}: This quantifies the correlation between a video's rank in the recommendation list and its utility score observed by the sock puppet across all recommendations. A stronger positive correlation indicates that the highest-ranked recommendations were more closely aligned with the sock puppet's preferences.
\end{itemize}

\noindent After analyzing the prevalence and prominence of utility, we examine the range and distribution of utility scores to identify any shifts and to contextualize the reinforcement dynamics

\para{Comparing reinforcement as prevalence of utility with the baseline.}  
To test our first sub-hypothesis (H1.1), we compared the mean utility of personalized recommendations for preference-aligned sock puppets ($U_p^s$) with that of control sock puppets ($U_c^s$).  
We collect all recommendations exposed to both preference-aligned and control sock puppets and evaluate their utility scores for the assigned preference. To compare the two distributions of utility scores, we perform a Kolmogorov-Smirnov (KS) test to assess the statistical significance of their differences and Cohen's d to measure the effect size of utility differences. A significantly higher mean utility for \( U_p^s \) compared to \( U_c^s \) would support H1.1, indicating that preference-aligned sock puppets experience greater reinforcement.
To further investigate the reinforcement effect, we analyzed the interquartile range (IQR) of utility scores for both preference-aligned and control sock puppets. Differences in the IQR provide insight into the overlap and divergence between the two groups. For instance, a narrower IQR for preference-aligned sock puppets, along with a higher mean utility, would suggest more concentrated reinforcement around emotionally aligned content. Conversely, a broader IQR with significant overlap may indicate a weaker or less consistent reinforcement effect.

\para{Comparing reinforcement as prominence of utility with the baseline.}  
To validate the second sub-hypothesis (H1.2), we examined the prominence of utility by analyzing the correlation between recommendation rank and utility scores for both preference-aligned and control sock puppets.
Specifically, we calculated the Spearman rank correlation coefficient to measure how prominently higher-utility videos were ranked within all recommendation traces for both preference-aligned and control sock puppets
A stronger significant and positive correlation for preference-aligned sock puppets compared to control sock puppets would validate our hypothesis and suggest that the algorithm learned the users preferences and prioritized higher-utility content.

\subsection{Results}\label{sec:h1:results}
In this section, we present the findings of our investigation into whether YouTube's recommendation algorithm reinforces meaningful emotional preferences. Overall, we observe significantly higher reinforcement, both in utility prevalence and utility prominence, for preference-revealing sock puppets ($U_{p}^{s}$) compared to the baseline ($U_{c}^{s}$). The statistical results supporting these findings are summarized in \Cref{tab:h1.1:results}.

\begin{table}[h!]
\small
\begin{tabular}{lrrrrr}
\toprule
\textbf{} & \textbf{Mean \%} & \textbf{Cohen's} & \textbf{IQR} & \textbf{Corr.} & \textbf{Corr.} \\
\textbf{Preference} & \textbf{Diff} & \textbf{d} & \textbf{Diff} & \textbf{(T)} & \textbf{(C)} \\
\midrule
\multicolumn{6}{l}{\textbf{News}} \\
Anger     & 138* & 0.73 & 0.33 & 0.05*  & -0.02 \\
Grievance & 60  & 0.37 & 0.16 & 0.02   & 0.01  \\
Group Identity & 84*  & 0.83 & 0.74 & 0.04*  & -0.01 \\
Negative  & 77*  & 1.03 & 0.01 & 0.01   & -0.01 \\
Positive  & 49*  & 0.50 & 0.14 & 0.01   & -0.01 \\
H-Frequency & 32  & 0.34 & 0.02 & 0.03*  & -0.01 \\
\midrule
\multicolumn{6}{l}{\textbf{Fitness}} \\
Anger     & 271* & 1.11 & 0.60 & 0.07*  & -0.01 \\
Grievance & 163* & 0.61 & 0.29 & 0.04*  & 0.02  \\
Group Identity & 74*  & 0.77 & 0.23 & 0.13*  & 0.03* \\
Negative  & 46*  & 0.70 & 0.00 & 0.07*  & 0.01  \\
Positive  & 41*  & 0.50 & 0.07 & 0.07*  & -0.03* \\
H-Frequency & 27  & 0.29 & 0.01 & 0.02   & -0.01 \\
\midrule
\multicolumn{6}{l}{\textbf{Gaming}} \\
Anger     & 116* & 0.80 & 0.34 & 0.08*  & 0.01  \\
Grievance & 148* & 0.63 & 0.24 & 0.06*  & 0.01  \\
Group Identity & 65*  & 0.81 & 1.06 & 0.10*  & -0.01 \\
Negative  & 55*  & 0.83 & 0.01 & 0.05*  & -0.00 \\
Positive  & 4   & 0.06 & 0.14 & 0.03*  & -0.01 \\
H-Frequency & 25  & 0.26 & 0.05 & 0.01   & 0.01  \\
\midrule
\multicolumn{6}{l}{\textbf{Random}} \\
Anger     & 90*  & 0.48 & 0.42 & 0.04*  & -0.00 \\
Grievance & 61  & 0.33 & 0.15 & 0.03*  & -0.01 \\
Group Identity & 55*  & 0.53 & 1.24 & 0.07*  & -0.01 \\
Negative  & 24  & 0.30 & 0.02 & 0.05*  & -0.02 \\
Positive  & 46*  & 0.53 & 0.10 & 0.04*  & -0.01 \\
H-Frequency & 20*  & 0.48 & -0.01 & 0.01   & 0.00  \\
\bottomrule
\end{tabular}
\caption{Statistical differences in utility between preference-selecting (Treatment) and random-selecting (Control) sock puppets across preferences and domains. \textit{Mean \% Diff} indicates the percentage difference in mean utility scores. \textit{Cohen’s $d$} measures the effect size of utility differences. \textit{IQR Diff} captures differences in utility score distributions. \textit{Corr. (T)} and \textit{Corr. (C)} represent the correlation between video rank and utility for treatment and control recommendations, respectively. Asterisks (*) indicate statistical significance (\(p < 0.05\)).}
\label{tab:h1.1:results}
\end{table}

\subsubsection{Up Next recommendations curated for preference-revealing sock puppets for videos selected based on preferences exhibit higher-than-baseline alignment with revealed preferences.} 
We conducted our analysis across four broad content categories: \textit{News}, \textit{Gaming}, \textit{Fitness}, and \textit{Random}.
Overall, utility scores were consistently higher for preference-revealing sock puppets, indicating significant reinforcement of revealed emotional preferences.  

\para{Higher average utility for preference-revealing sock puppets.}  
Sock puppets that revealed their preferences consistently received recommendations with higher utility scores compared to the control sock puppets.
Notably, \textit{anger} demonstrated substantial reinforcement across all categories: 138\% higher utility in \textit{News}, 271\% in \textit{Fitness}, 116\% in \textit{Gaming}, and 90\% in \textit{Random}.
Similar patterns emerged for preferences related to \textit{negative} sentiment, \textit{grievances}, and \textit{group identification}, with meaningful increases in utility across domains.  
Looking more closely, reinforcement varied depending on the preference and seed category.
For example, \textit{grievance} preferences were strongly amplified in the \textit{Fitness} domain (148\%), while \textit{group identification} preferences showed substantial reinforcement in the \textit{News} domain (84\%).
In contrast, notably, arbitrary or nonsensical preferences, such as selecting videos based on letter frequency (\textit{H-Frequency}), did not exhibit significant utility gains, except in random selections.
Similarly, preferences for \textit{positive} sentiment showed limited or no reinforcement in categories like \textit{Fitness} and \textit{News}.  

\para{Systematic shifts in distribution of utility for preference revealing sock puppets.}
The IQR differences in \Cref{tab:h1.1:results} indicate how the distributions of utility scores shifted in response to revealed preferences.
For preferences like \textit{group identity}, large IQR increases (e.g., 1.06 in \textit{gaming}) suggest that reinforcement comes not just from a uniform upward shift, but also from expanding the upper tail, highlighting a subset of particularly high-utility videos.
In contrast, preferences such as \textit{negative} in \textit{news} show high mean increases with almost no IQR change, implying a more even elevation of the entire distribution rather than selective tail growth.
Overall, some preferences trigger selective amplification at the high end of the utility spectrum, while others uniformly raise the overall utility level without reshaping the distribution's tails.

\begin{figure*}[h]
    \centering
    \includegraphics[width=1\textwidth]{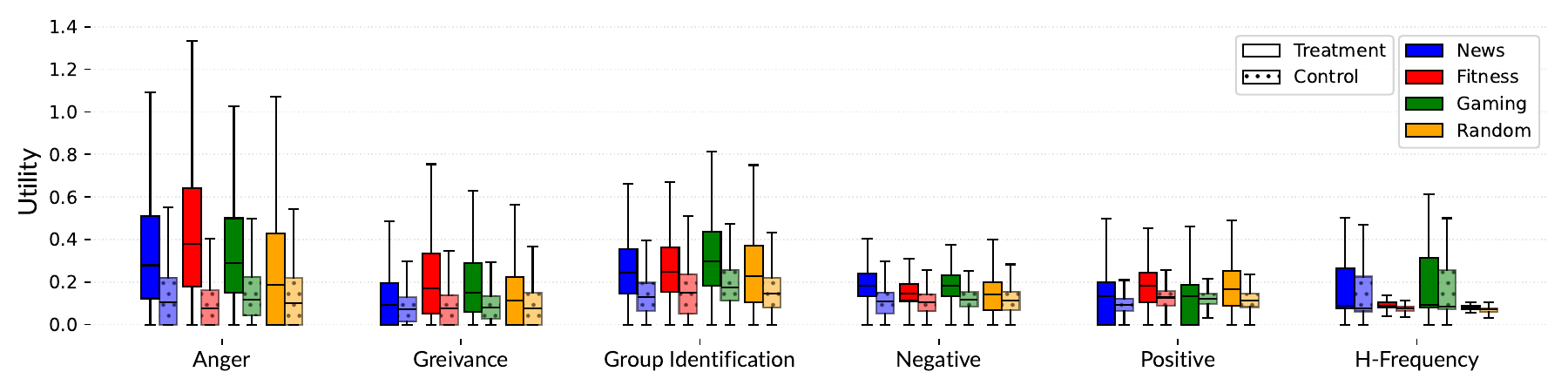}
    \caption{Distribution of utility values derived from Up Next recommendations for preference-aligned (treatment) and control sock puppets, grouped by seed domain. Note that some utility values have been scaled independently better visual representation within the graph. For exact values refer to \cref{tab:h1.1:results}.}
    \label{fig:h1.1:box}
\end{figure*}

\subsubsection{Up Next recommendations for selected videos exhibit higher prominence of alignment of preferences compared with baseline} 

A strong positive correlation emerged between video utility scores and their ranking positions for preference-aligned sock puppets, indicating that higher-ranked recommendations align more closely with revealed emotional preferences.
For example, Anger preferences consistently exhibited significant correlations between utility and rank across multiple categories, including News (\(r = 0.05\), \(p < 0.05\)) and Fitness (\(r = 0.07\), \(p < 0.05\)).
Although control sock puppets also exhibited a positive rank-utility correlation, it was notably weaker.
For instance, in the case of \textit{group identification}, the rank-utility correlation for preference-revealing sock puppets was \(r = 0.04\) in News and \(r = 0.13\) in Fitness, compared to negligible or slightly negative correlations for the control group (\(-0.01\) and \(0.03\), respectively). 
In contrast, preferences for arbitrary selection criteria, such as H-Frequency, showed minimal or negligible correlations between utility and rank, such as \(r = 0.03\) in News and \(r = 0.01\) in Fitness, further indicating the algorithm's inability to reinforce nonsensical preferences.

\subsection{Takeaways}  
Our results validate both sub-hypotheses and, consequently, the main hypothesis that YouTube reinforces emotional preferences through increased prevalence and prominence of preference-aligned content. These findings can be synthesized into three key takeaways:

\begin{enumerate}
    \item \textbf{YouTube reinforces meaningful preferences, implying their recognition.}  
    Preference-revealing sock puppets were consistently exposed to significantly higher-utility content compared to their control counterparts.
    Reinforcement was particularly strong for preferences associated with negative emotions, such as anger, grievance, and group identification.
    However, reinforcement was not universal: preferences with limited representation in the broader user base were less likely to exhibit this effect.
    These results highlight the importance of collective user behavior in the algorithm's training data—since preferences with insufficient representation in the broader user base may not be actively reinforced.
    More specifically, this suggests the algorithm \textit{recognizes} assigned preferences as revealed preferences.

    \item \textbf{Reinforcement can result from a uniform shift or skewness in utility.}  
    For many preferences, reinforcement manifests as an overall upward shift in the utility distribution rather than a change in skewness.
    This indicates, for these cases, the algorithm reduces access to moderate- or low-utility recommendations, creating an “emotional filter bubble” effect where users are increasingly exposed to emotionally aligned content. 
    Alternatively, for preferences where reinforcement manifests as the inclusion of a subset of high-utility videos, this suggests a selective amplification of the upper tail of the distribution.
    In these cases, the algorithm disproportionately surfaces videos that align strongly with revealed preferences.
    \item \textbf{Reinforcement is reflected in video rankings.}  
    The algorithm prioritizes preference-aligned content by ranking videos with higher utility scores more prominently.
    While factors such as relevance, popularity, and freshness contribute to video rankings, our findings reveal that personalization based on revealed preferences plays a significant role.
    This is confirmed by the weak or negligible correlation between rank and utility for control sock puppets, i.e. the reinforcement is not a byproduct of domain-level influences but rather stems from the algorithm's personalization mechanisms.
\end{enumerate}

These results demonstrate that YouTube's recommendation algorithm reinforces emotional preferences, increasing both the prevalence and prominence of preference-aligned content. This dual reinforcement, prominence and prevalence, shapes user experiences and contributes to a more emotionally reinforcing environment.
\section{Personalized recommendations are more reinforcing than contextual recommendations}  
In this section, we compare reinforcement of emotional preferences between personalized and contextual recommendations.
\textit{Contextual recommendations} are generated without access to a user's watch history, relying solely on the context of the currently selected video.
In contrast, \textit{personalized recommendations} incorporate learned user preferences through the user's watch history, while also considering contextual factors and collaborative filtering based on collective user behavior.
Because they can leverage watch history to learn users' revealed preferences, we hypothesize that personalized recommendations exhibit a higher degree of reinforcement compared to contextual recommendations.

If validated, this would suggest that YouTube's recommendation algorithm amplifies the reinforcement of users' emotional preferences through personalization, exceeding the reinforcement achieved by contextual recommendations alone.
To test our hypothesis, we compare the reinforcement within personalized Up Next recommendations to the reinforcement within contextual recommendations for the same set of preference-aligned videos.
Here, similar to \textit{H1}, we operationalize reinforcement as prevalence and prominence of utility within the recommendations (see \cref{sec:h1:methods}.
\subsection{Methods}
In this section, we outline the methods used to obtain contextual recommendations and compare their reinforcement with the personalized recommendations curated for preference-revealing sock puppet.

\para{Collecting contextual and personalized recommendations.}
In addition to collecting all exposed personalized recommendations during the experiment, we also recorded contextual recommendations for each selected video, as described in \Cref{sec:data:record-aligned}.
To obtain contextual recommendations, we request recommendations for the selected video \( v_i \) from a sock puppet that is not signed in to the platform.
This allowed us to capture both personalized recommendations (\( R_{personalized}(v_i) \)) and contextual recommendations (\( R_{contextual}(v_i) \)) for the same video \( v_i \).

\para{Comparing reinforcement of preferences.}
Subsequently, we measure the reinforcement of preferences from both personalized and contextual recommendations using the methods outlined in \Cref{sec:h1:methods}.
We compare the reinforcement of preferences between these two types of recommendations, focusing on three key metrics: the change in mean utility, the skewness of utility distribution, and the prominence of utility.
This comparison enables us to test our hypothesis regarding the distinct role of personalization in reinforcing user preferences beyond what is achieved through contextual recommendations alone.

\subsection{Results}
Overall, our findings reject the hypothesis that personalized recommendations exhibit significantly higher reinforcement compared to contextual recommendations.
Instead, our results support the alternative hypothesis, suggesting that \textit{contextual recommendations are more reinforcing than personalized recommendations}. We provide a summary of our results in \cref{tab:h2.1:results}

\begin{table}[ht!]
\small
\begin{tabular}{lrrrrr}
\toprule
\textbf{} & \textbf{Mean \%} & \textbf{Cohen's} & \textbf{IQR} & \textbf{Corr.} & \textbf{Corr.} \\
\textbf{Preference} & \textbf{Diff} & \textbf{d} & \textbf{Diff} & \textbf{Persl.} & \textbf{Ctxl.} \\
\midrule
\multicolumn{6}{l}{\textbf{News}} \\
Anger       & -7.94*  & -0.18 & -0.06 & 0.05*  & 0.08*  \\
Grievance         & 12.26*  & 0.24  & 0.01  & 0.02   & 0.08*  \\
Group Identity     & -4.59*  & -0.18 & 0.08  & 0.04*  & 0.10*  \\
Negative    & 3.59*   & 0.17  & 0.00  & 0.01   & 0.07*  \\
Positive    & -3.77   & -0.10 & -0.01 & 0.01   & 0.01   \\
H-Frequency    & -1.22   & -0.02 & -0.00 & 0.03*  & 0.07*  \\
\midrule
\multicolumn{6}{l}{\textbf{Fitness}} \\
Anger       & -5.89*  & -0.15 & -0.07 & 0.07*  & 0.10*  \\
Grievance         & -9.87*  & -0.19 & -0.02 & 0.04*  & 0.05*  \\
Group Identity     & -4.88*  & -0.19 & -0.20 & 0.12*  & 0.12*  \\
Negative    & -5.18*  & -0.25 & 0.00  & 0.07*  & 0.07*  \\
Positive    & 0.05    & 0.00  & -0.00 & 0.07*  & 0.08*  \\
H-Frequency    & -1.78   & -0.03 & -0.01 & 0.02*  & 0.05*  \\
\midrule
\multicolumn{6}{l}{\textbf{Gaming}} \\
Anger       & -9.16*  & -0.25 & -0.11 & 0.08*  & 0.09*  \\
Grievance         & -9.62*  & -0.14 & -0.04 & 0.06*  & 0.06*  \\
Group Identity     & -7.81*  & -0.34 & -0.25 & 0.10*  & 0.12*  \\
Negative    & -0.82   & -0.04 & -0.00 & 0.05*  & 0.10*  \\
Positive    & -8.71*  & -0.27 & -0.01 & 0.03*  & 0.07*  \\
H-Frequency    & -5.23   & -0.08 & -0.03 & 0.01   & 0.02   \\
\midrule
\multicolumn{6}{l}{\textbf{Random}} \\
Anger       & -7.59*  & -0.19 & -0.08 & 0.06*  & 0.09*  \\
Grievance         & -7.14*  & -0.11 & -0.04 & 0.05*  & 0.06*  \\
Group Identity    & -5.90*  & -0.23 & -0.22 & 0.09*  & 0.11*  \\
Negative    & -0.64   & -0.03 & 0.00  & 0.04*  & 0.08*  \\
Positive    & -3.35*  & -0.09 & -0.00 & 0.04*  & 0.05*  \\
H-Frequency    & -2.82   & -0.04 & -0.01 & 0.02*  & 0.05*  \\
\bottomrule
\end{tabular}
\caption{Statistical differences between personalized and contextual recommendations across preferences and content categories. \textit{Mean \% Diff} indicates the percentage difference in mean utility scores, with negative values showing higher utility for contextual recommendations. \textit{Cohen’s $d$} measures the effect size of utility differences. \textit{IQR Diff} captures differences in utility score spread. \textit{Corr. Persl} and \textit{Corr. Ctxl} represent the correlation between video rank and utility for personalized and contextual recommendations, respectively. Asterisks (*) denote statistical significance ($p < 0.05$)}
\label{tab:h2.1:results}
\end{table}

\begin{figure}[t]
    \centering
    \includegraphics[width=0.5\textwidth]{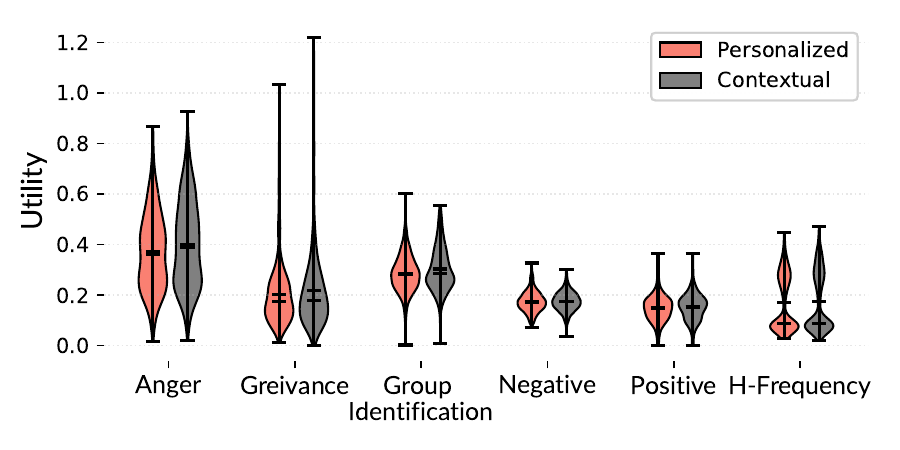}
    \caption{Distribution of utility values  from personalized and contextual Up Next recommendations. We aggregate recommendations across all domains. Note that some utility values have been scaled independently better visual representation within the graph. For exact values refer to \cref{tab:h2.1:results}.}
    \label{fig:h2.1:box}
\end{figure}

\subsubsection{Contextual recommendations curated for preference-revealing sock puppets for videos selected based on preferences exhibit higher reinforcement than personalized recommendations.} 
Our findings, summarized in \Cref{tab:h2.1:results}, demonstrate contextual recommendations to exhibit higher prevalence and prominence of utility compared to its personalized counterpart.

\para{Contextual recommendations have higher prevalence of utility.}
Across all content categories, contextual recommendations consistently demonstrated higher mean utility scores than personalized recommendations.
This trend was observed across all tested emotional preferences, with the exception to a few combinations of preference and seed domain.
For example, in the case of anger-related preferences, contextual recommendations achieved significantly higher mean alignment than personalized recommendations across all domains, including \textit{News} (-7.94\%), \textit{Fitness} (-5.89\%), and \textit{Gaming} (-9.16\%). 
Similar patterns were observed for preferences associated with sadness, group identification, and negativity, where contextual recommendations consistently outperformed their personalized counterparts.
Preferences such as positivity and arbitrary criteria (i.e., H-Frequency) displayed minimal differences between personalized and contextual recommendations.
Additionally, the distributions of utility scores between the two types of recommendations did not exhibit large differences, with utilities within personalized recommendations yielding a slightly narrower distribution.

\para{Contextual recommendations have higher prominence of utility.}  
The correlation between utility scores and the rank of recommended videos in the Up Next list demonstrates consistent differences between contextual and personalized recommendations.
Contextual recommendations exhibited stronger positive correlations between utility and rank, indicating that highly ranked videos were more aligned with the sock puppets' preferences.
For instance, in the case of the anger preference, the correlation for contextual recommendations in \textit{News} (Corr. \(C = 0.08\)) was higher than that for personalized recommendations (Corr. \(P = 0.05\)).

\subsection{Takeaways}
Our results fail to validate our hypothesis. Instead, they support an alternative hypothesis: for the same video selected based on its preference, contextual recommendations are more reinforcing than personalized recommendations on YouTube.

\begin{itemize}
    \item \textbf{Contextual recommendations reinforce revealed preferences.} Contextual recommendations rely heavily on information present within the context—which is limited to the currently watched video
    This outcome may stem from collaborative filtering capturing collective behavior, resulting in contextual recommendations that reflect broader collective user behavior, which could inadvertently align with emotional preferences.
    Alternative explanations may include the algorithm's reliance on freshness and popularity when curating recommendations, factors that could incidentally correlate with certain emotional preferences.
    
    \item \textbf{Contextual recommendations reinforce revealed preferences more than personalized recommendations.}
    Personalized recommendations theoretically have access to all information available to contextual recommendations in addition to the history of users' revealed preferences.
    In our experiment, this history is explicitly limited to the assigned emotional preference. Yet, we observe the inclusion of this history dampens the reinforcement of user preferences when compared with reinforcement within contextual recommendations.
    This surprising behavior can be explained by the possibility that personalization mechanisms may prioritize objectives post-hoc such as content diversity, novelty, or user retention over strict preference alignment. 
\end{itemize}
\section{YouTube amplifies exposure to emotionally aligned content.}
In this section, we investigate whether users' exposure to reinforcing preferences is amplified.
In prior literature, amplification is commonly used to imply an intensification of user biases beyond reinforcement and what can be directly predicted from their immediate behavior.
In our context, while reinforcement implied a process of user revealing their preferences and the algorithm sustaining the preferences, amplification refers to the change in reinforcement over time and over different contexts.
%
%
We operationalize our main amplification hypothesis using two separate sub-hypotheses addressing the \textit{intensification} over time and the \textit{persistence} of reinforcement.
\begin{enumerate}[label=\textbf{H3.\arabic*.}, ref=H3.\arabic*]
    \item \textit{Up Next recommendations become increasingly more reinforcing as the user continues to engage with preference-aligned videos.} 
    Throughout the experiment, we measure the progression of reinforcement observed by the sock puppets over time, focusing on the gradient of this reinforcement. This analysis aims to answer: does the reinforcement increase as engagement continues, indicating a growing intensity in reinforcement?
    \item \textit{Up Next recommendations for pre-defined videos, not selected for their emotional alignment, exhibit higher-than-baseline alignment with the user's revealed preferences.}
    Once the sock puppet preferences are revealed, we evaluate the emotional reinforcement within recommendations for a predefined set of videos.
    This approach isolates the effect of personalization, enabling us to assess whether recommendations for unrelated videos also exhibit a reinforcement effect driven by the user's revealed preferences.
\end{enumerate}

If our sub-hypotheses are valid, they suggest that YouTube's recommendation algorithm progressively serves users content that becomes increasingly aligned with their emotional preferences, both within preference-aligned contexts and in broader, unrelated contexts.
Such outcomes would have implications on how personalization shapes a user's online experience of exposure to preference aligned content. 

\subsection{Method}
We characterize amplification as two different phenomena, studied separately by the two sub-hypotheses: 1) increase in reinforcement over time,
and 2) reinforcement in the recommendations of unrelated pre-defined videos.

\para{Measuring intensification of reinforcement through a moderation analysis.}
Using the collected Up Next recommendations from both the preference-aligned (treatment) and control sock puppets, we investigate whether the reinforcement of preferences increases over time for the treatment group compared to the control.
Here, time represents the sequential steps in our experiment, i.e., the depth of the recommendation path, corresponding to the number of instances where preferences are revealed.
We employ a moderation analysis to analyze whether the reinforcement of preferences (dependent variable) varies based on recommendation depth (independent variable) and experimental condition (moderator: treatment or control).
Moderation analysis tests whether the relationship between an independent variable and a dependent variable depends on the value of a third variable: the moderator.
In our analysis, the dependent variable is the normalized mean reinforcement at each depth, calculated as prevalence of utility.
The independent variable is the depth of the recommendation trace, which reflects the extent of the sock puppet's engagement history and how much information the algorithm has accumulated about its behavior.
Our primary moderator is the experiment condition: treatment (preference-aligned selections) or control (random selection).
We also include another moderator representing the starting point for the sock puppet, i.e., one of four domains from which the seed is selected: \textit{News}, \textit{Fitness}, \textit{Gaming}, \textit{Random}.

Representing the moderation analysis as an equation excluding the seed domain moderators: 

\[
R_{ij} = \beta_0 + \beta_1 \text{Depth}_i + \beta_2 T_j + \beta_3 (\text{Depth}_i \times T_j) + \epsilon_{ij}
\]

\noindent
Where \( R_{ij} \) is the normalized mean reinforcement for recommendation depth \( i \) and treatment condition \( j \); \( \text{Depth}_i \) represents the step in the recommendation sequence; \( T_j \) denotes the experimental condition (1 for treatment, 0 for control); \( \text{Depth}_i \times T_j \) captures the interaction between depth and treatment; \( \epsilon_{ij} \) is the residual error. The coefficients \( \beta_1 \), \( \beta_2 \), and \( \beta_3 \) represent the direct effect of depth, the effect of treatment, and the moderation effect, respectively.



We accept the hypothesis if the moderation effect, represented by \(\beta_3\), is positive and statistically significant (\(p < 0.05\)).
This would indicate that reinforcement increases more rapidly with depth in the treatment group compared to the control group.

\para{Measuring persistence of reinforcement on pre-defined videos.}
In this section, we evaluate reinforcement within personalized recommendations for a predefined set of videos that are randomly selected from the domains sets (see \cref{sec:data:record-predefined}.
We measure and compare the reinforcement—prevalence and prominence of utility—for recommendations served to preference-revealing and control sock puppets using similar methods as defined in \Cref{sec:h1:methods}. 
Observing significantly higher reinforcement for preference-revealing sock puppets compared to random-selecting sock puppets would indicate a persistent reinforcement, i.e., amplification of reinforcement that is beyond what is predicted by the users' immediate behavior, driven solely by personalization—hence validating our hypothesis.

\subsection{Results}
Overall, our findings validate our main hypothesis that personalized recommendations amplify reinforcement over time and beyond immediate context.

\subsubsection{Up Next recommendations get increasingly more reinforcing.} Our findings demonstrate that personalized recommendations become increasingly reinforcing, compared to the baseline, as the sock puppet continues revealing its preferences and selecting videos that align with those preferences.
The findings from our moderation analysis, in summary, indicate that while the reinforcement increased for all conditions (treatment and control) as the sock puppets went deeper into the recommendations, the treatment sock puppets experienced significantly higher reinforcement for most preferences and seeds.
We share our findings in \cref{tab:h3.2:results}.

\para{Reinforcement increases with depth.} 
Our moderation analysis finds that reinforcement generally increases with depth, particularly for certain emotional preferences.
This indicates that as sock puppets, treatment or control, starting from a seed video, traverse down the recommendations, the recommendations become more emotionally aligned.
For instance, the coefficient for \textit{Depth} ($\beta_{\textit{Depth}}$) is positive for preferences such as \textit{Anger} ($0.002$) and \textit{Negative} ($0.004$), indicating that reinforcement increases.
For other preferences like \textit{Grievance}, \textit{Group Identity}, and \textit{Positive}, the depth coefficient remains close to zero, suggesting minimal direct change in reinforcement over time.

\para{The increase in reinforcement is significantly higher for preference-assigned sock puppets.} Overall, our analysis demonstrates that the increase in reinforcement is significantly higher for preference-assigned sock puppets compared to the baseline.
Beginning with the coefficients for \textit{Treatment} ($\beta_{\textit{T}}$), these values represent the initial reinforcement after the first selected video based on the assigned preference.
These coefficients are large and positive across all emotional preferences, such as \textit{Anger} ($0.925$), \textit{Group Identity} ($1.040$), and \textit{Negative} ($0.913$).
Furthermore, the significant interaction effects between depth and treatment ($\beta_{\textit{D} \times \textit{T}}$) reinforce this trend.
For example, the interaction effect for \textit{Anger} ($0.005^*$) and \textit{Grievance} ($0.004^*$) confirms that reinforcement increases at a faster rate for preference-assigned sock puppets compared to the control group, which selects videos randomly.

\para{The increase in reinforcement is shaped by the seeds.}   Finally, we evaluate the role of the starting seed category and its influence on the reinforcement.
We find that the rate of reinforcement varies depending on the seed, as evidenced by the seed-specific interaction terms.
For example, \textit{Fitness} seeds exhibit positive interaction effects for preferences such as \textit{Anger} ($0.009$) and \textit{Positive} ($0.007$), indicating a faster increase in reinforcement when starting from fitness-related content.
In contrast, \textit{Random} seeds tend to show weaker or negative interaction effects, suggesting a slower rate of reinforcement.
Similarly, \textit{Gaming} and \textit{News} seeds display mixed effects depending on the preference.

\begin{table}[h!]
\small
\centering

\label{tab:reinforcement_metrics}
\begin{tabular}{lccc}
\toprule
& \textbf{Mean \%}           & $\beta$ & $\beta$ \\
\textbf{Preference} & \textbf{Diff}           & \textbf{Treatment.} & \textbf{Control.} \\
\midrule
\multicolumn{4}{l}{\textbf{News}} \\
Anger        & 14.25*                   & 0.021*                  & 0.001                 \\
Grievance    & 0.05                     & 0.002                   & 0.003                 \\
Group Identity & 13.35*                 & 0.032                   & -0.037*               \\
Negativity   & 6.15*                    & 0.002*                  & 0.002                 \\
Positivity   & 5.95*                    & 0.001                   & -0.001                \\
H-Freq.      & 1.59                     & 0.002                   & 0.002                 \\
\midrule
\multicolumn{4}{l}{\textbf{Gaming}} \\
Anger        & 5.60                     & 0.003                   & -0.001                \\
Grievance    & -5.03                    & -0.001                  & -0.001                \\
Group Identity & 10.07*                 & 0.057                   & -0.016                \\
Negativity   & 11.87*                   & 0.002*                  & -0.000                \\
Positivity   & -5.36                    & -0.001                  & -0.001*               \\
H-Freq.      & 2.07                     & 0.003*                  & -0.000                \\
\midrule
\multicolumn{4}{l}{\textbf{Random}} \\
Anger        & 7.29                     & 0.029*                  & 0.024*                \\
Grievance    & 6.20                     & 0.007                   & 0.006*                \\
Group Identity & 15.48*                 & 0.031                   & -0.025                \\
Negativity   & 5.56*                    & 0.003*                  & 0.003*                \\
Positivity   & 13.19*                   & 0.004                   & -0.001                \\
H-Freq.      & 2.13                    & 0.000                   & 0.001                 \\
\bottomrule
\end{tabular}
\caption{Reinforcement metrics for emotional preferences across \textit{News}, \textit{Gaming}, and \textit{Random} domains for the predefined set of videos. 
The "Mean \% Diff" column represents the average percentage difference in alignment with emotional preferences compared to baseline recommendations. The \(\beta_{pers}\), change in reinforcement over time for treatment sock puppets,  reflects the regression coefficient for the alignment trend in personalized recommendations. The \(\beta_{ctx}\) denotes the regression coefficient for control sock puppet. Significant values ($p<0.05$) are marked with an asterisk (*).}
\label{tab:h3.2:results}
\end{table}
\begin{table}[ht!]
\centering
\label{tab:moderation_results}
\resizebox{0.5\textwidth}{!}{
\begin{tabular}{lrrrrrrrr}
\toprule
\textbf{Measure} & $R^2$ & $\beta_{\textit{Depth}}$ & $\beta_{\textit{T}}$ & $\beta_{\textit{D} \times \textit{T}}$ & $\beta_{\textit{Fitness}}$ & $\beta_{\textit{Gaming}}$ & $\beta_{\textit{News}}$ & $\beta_{\textit{Random}}$ \\ 
\midrule
\textbf{Anger}      & 0.391 & 0.002  & 0.925  & 0.005*  & 0.009  & 0.000  & -0.001  & -0.003 \\
\textbf{Grievance}        & 0.297 & -0.000 & 0.699  & 0.004* & 0.007  & 0.007  & -0.006  & -0.003 \\
\textbf{Group Id.}    & 0.441 & -0.001 & 1.040  & 0.004*  & 0.002  & 0.003  & 0.002   & -0.004 \\
\textbf{Negative} & 0.333 & 0.004  & 0.913  & 0.002    & -0.004 & 0.002  & 0.007   & -0.004 \\
\textbf{Positive} & 0.345 & -0.000 & 0.434  & 0.006*  & 0.007  & -0.010 & -0.001  & 0.008  \\
\textbf{H-Freq.}  & 0.260 & 0.003  & 0.233  & 0.002    & -0.000 & 0.002  & 0.003   & -0.003 \\
\bottomrule
\end{tabular}
}
\caption{Moderation analysis results for reinforcement across measures. The table reports the proportion of variance explained (\(R^2\)) and the coefficients for key variables: \(\beta_{\textit{Depth}}\) (effect of recommendation depth), \(\beta_{\textit{T}}\) (baseline treatment effect), and \(\beta_{\textit{D} \times \textit{T}}\) (interaction of depth and treatment). The seed-specific interaction terms (\(\beta_{\textit{Fitness}}\), \(\beta_{\textit{Gaming}}\), \(\beta_{\textit{News}}\), \(\beta_{\textit{Random}}\)) represent how reinforcement trends vary across different starting content categories. Significant effects (\(p < 0.01\)) are marked with an asterisk (*).}
\label{tab:h3.1:results}
\end{table}

\begin{figure}[h]
    \centering
    \includegraphics[width=0.45\textwidth]{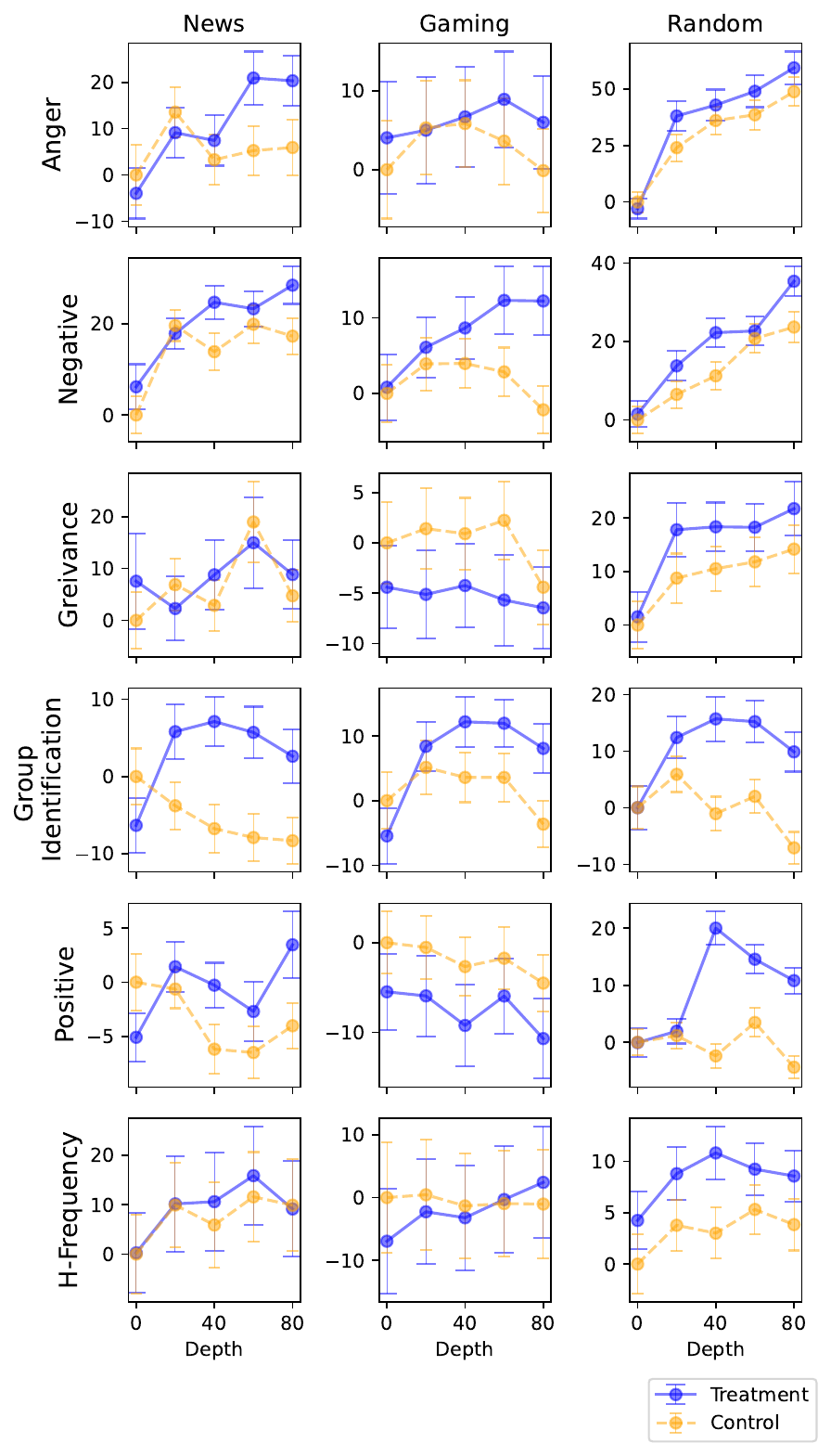}
    \label{fig:h2.1:box}
    \caption{Mean utility within Up Next recommendations of predefined videos for preference-revealing (treatment) and random-selecting (control) sock puppets. The values are shown as the percentage difference from the first control-observed recommendations. All values at the 0\textsuperscript{th}
 essentially show utility within contextual recommendations.}
\end{figure}

\subsubsection{Reinforcement persists across recommendations for predefined videos not selected to optimize preferences.}
We analyzed the reinforcement of emotional preferences within recommendations for predefined videos.
The results reveal consistent patterns of emotional reinforcement greater than the baseline as characterized by the counterfactual sock puppets. 
For example, \textit{anger} showed significant reinforcement in \textit{News} videos, with a 14.25\% higher alignment compared to baseline. Gaming and random categories also demonstrated increases (5.60\% and 7.29\%, respectively), although with less pronounced effects. 
By contrast, positivity displayed substantial reinforcement for random seeds (13.19\%, \textit{d} = 0.51), while news and gaming showed minimal or negative effects.
These patterns indicate a stronger focus on amplifying emotionally charged content, such as negativity and group identification, over uplifting or neutral content.
Notably, h-frequency, a meaningless control preference, showed little to no reinforcement across categories, confirming the algorithm’s focus on meaningful emotional patterns.

\subsubsection{Reinforcement intensifies within recommendations for predefined videos not selected to optimize preferences.}
We analyze the regression coefficients of personalized recommendations (\(\beta_{pers}\)) and contextual recommendations (\(\beta_{ctx}\)) to evaluate whether reinforcement intensifies over time.
As shown in \Cref{tab:reinforcement_metrics}, we observe a consistent pattern of increasing alignment with selected emotionally charged preferences, particularly for \textit{anger}, \textit{negativity}, and \textit{group identity}, especially when compared to recommendations shown to the control sock puppet.
By contrast, preferences such as \textit{positivity} exhibit flat or negative trends in both treatment and control recommendations.

\subsection{Takeaways}
Our results provide evidence for amplification of emotional reinforcement through two different methods.
We find the reinforcement to persist in future recommendations for videos unrelated to the preference, and the intensification of reinforcement over time. Together, our findings establish the following key takeaways. 

\begin{itemize}
    \item \textbf{Intensification of reinforcement through personalization, preference hunting, and random walking.}
    Our findings indicate intensity of reinforcement to be dependent on the depth and the condition of the sock puppet. This suggests a positive increase in the reinforcement of revealed emotional preferences.
    More specifically, an increase in depth is associated with increased reinforcement within our experiment's context. 
    Interestingly, we also find sock puppets randomly selecting videos from a seed occasionally under certain conditions observe an increase in intensification of reinforcement.
    The phenomena of intensified reinforcement can be attributed to two interconnected explanations: (1) as depth increases, the system's confidence in the revealed preferences grows, and (2) as depth increases, the likelihood of selecting a video that strongly aligns with the preferences also increases. However, our analysis lacks the robustness to definitively separate these two explanations.
    \item \textbf{Persistent reinforcement of revealed preferences.}
    Our findings establish the presence of reinforcement even in videos that are unrelated to the seed videos.
    Where as reinforcement within recommendations for videos selected due to their high alignment with preference can be attributed due to it contextual information, reinforcement observed here isolates its attribution to personalization and specifically revealed preferences.
\end{itemize}

Taken together, our findings present a case for amplification through both difference recommendation surfaces and depths.
Results from our intensification imply a user-driven platform-assisted rabbit holing where the platform serves increasingly more aligned videos to the user.
Where as, persistence reinforcement presents a case for an emotional filter bubble effect.
\section{Discussion}

In this section, we outline the conclusions of our findings (\cref{sec:discussion:conclusions}) and their implications (\Cref{sec:discussion:implications}). We then use our findings to discuss the role of media platforms and content curation algorithms in shaping user behavior and beliefs \cref{sec:discussion:comments}, emphasizing how the optimization for engagement may subtly influence revealed preferences and constructed user choices. Finally, we discuss the limitations of our work \cref{sec:discussion:limitations}.

\subsection{Summary of our findings.}\label{sec:discussion:conclusions}
Overall, our findings provide experimental evidence that emotional revealed preferences are recognized and reinforced by YouTube's recommendation algorithm. 

\para{YouTube's recommendation algorithm reinforces emotional preferences.} More specifically, when users engage with content that matches a particular assigned emotion—especially a negative one—subsequent recommendations become increasingly aligned with that emotion. This effect was most pronounced for anger, grievance, and negativity, particularly in news contexts, where the overlap between recommended content and the baseline set of recommendations was very limited or nonexistent. Conversely, positive emotions—especially within gaming contexts—received the least reinforcement.

\para{Contextual recommendations are more reinforcing than personalized recommendations.} The contextual information alone from a single \textit{currently watching} preference-aligned video was responsible for a more reinforcing set of recommendations compared to personalized recommendations, which had access to a combination of contextual information and engagement history information. This effect was more pronounced for negative emotions (with the exception of grievance in news contexts) compared to positive emotions, and was absent for meaningless preferences.

\para{The reinforcement of preferences become intensified over time and is persistent across different unrelated contexts.} As the user engages with preference-aligned content, the average utility within the recommendations gradually increases compared to the baseline (measured using control sock puppets) at that depth. In addition to this, once preferences are revealed, recommendations in different contexts—i.e., recommendations for predefined videos not previously engaged with—persist to exhibit higher-than-baseline reinforcement.

\subsection{Implications of our findings.}  \label{sec:discussion:implications}
In this section, we explore the implications of YouTube’s emotional preference reinforcement, focusing on the influence of collective user behavior, how an emotional bubble effect emerge, and the possible consequences of sustaining negative emotions. 

\para{Collective user behavior reinforces emotional reinforcement.} YouTube's recommendations system relies on a form of collaborative filtering driven by deep neural networks \cite{covington_deep_2016}. Through this, the content is \textit{perceived} by the platforms as how users have engaged with it and similarly users are characterized by the content they have engaged with. Reinforcement of emotional preferences imply that there is a sufficiently large user base consistently engaging with such emotional content, enabling the algorithm to `recognize' and reinforce it. The absence of reinforcement for a `meaningless' preference further supports this explanation. This demonstrates the role played by, not the algorithm, but rather and the collective behavior of the user base. Additionally, as contextual recommendations are driven purely by collective user behavior, higher reinforcement in contextual recommendations compared to personalized recommendations once again highlight how collective user behavior can shape individual recommendation paths.

\para{Up Next recommendations skew towards exploiting emotional preferences.}
A notable observation from our findings is the dramatic shift in recommendations once a specific emotional preference is revealed. In some instances, the overlap in interquartile ranges between baseline and preference-aligned recommendations was almost nonexistent, indicating a near-complete realignment of the recommendation set. This phenomenon suggests the emergence of a potential filter bubble effect, where users may become confined to a narrow segment of the content space, limiting both emotional diversity in their recommendations. This outcome highlights YouTube's tendency to skew the explore–exploit tradeoff heavily toward exploitation, as similarly evidenced in prior research \cite{habib_uncovering_2024}.

\para{Sustaining negative emotional content poses risks.} 
A key concern emerging from our study is the platform’s tendency to perpetuate negative emotions (e.g., anger, grievance).
While it is unsurprising that dramatic or provocative content often garners higher engagement \cite{brady_emotion_2017}, we find recommendation algorithm’s reinforcement can create feedback loops that keep users locked into negative emotional states.
Beyond the immediate user experience—where a constant feed of aggrieved or outraged content risks affecting well-being—there are broader implications for public discourse \cite{martel_reliance_2020}. Over-reinforcement of negative preferences may fuel polarized debates, amplify grievances within communities, contribute to echo chambers where only one emotional tone is dominant, and for the individual, exacerbate depression, anxiety and stress.

\subsection{Giving the user what they want.} \label{sec:discussion:comments}
Sock puppet studies, like ours, aim to understand the \textit{behavior} of algorithms under specific, controlled conditions. While they lack the external validity of observational or experimental studies due to their artificial nature, they offer valuable insights into algorithmic functioning. Our study, uses sock puppets with defined emotional preferences to examine how YouTube's recommendation algorithm reinforces these preferences. Notably, we found that the algorithm can indeed recognize and reinforce emotional preferences, which is a noteworthy finding in itself. However, a simplistic interpretation of our work might be that the algorithm effectively "gives users what they want". While accurate, we believe the implications of our study extend far beyond this observation. The optimization for engagement, particularly through watch time, raises several crucial questions regarding the subtle and potentially problematic ways recommendation algorithms shape user preferences and experiences. We highlight three key arguments below:

\para{Does watch time capture true user preferences?}
Historically, recommendation algorithms, including YouTube's, have relied on various metrics as proxies for user engagement. Earlier optimization for clickthrough rates incentivized sensationalist "clickbait" content, demonstrating how easily metrics can be gamed. The shift to watch time aimed to address this but presents its own limitations. In their article, Thorburn \etal \cite{thorburn_what_2022} argue that engagement metrics like watch time often conflate user choices with preferences and welfare, oversimplifying what users genuinely want. Watch time may reflect passive consumption driven by autoplay or addictive design rather than true satisfaction or intent. It prioritizes longer content, often at the expense of concise, purposeful videos, and amplifies emotionally manipulative content, creating feedback loops that distort user experiences and encourage polarization or unhealthy habits. Our experiment corroborate with these conjectures and provide evidence of how emotional preferences are considered engaging by the algorithm. To move beyond the pitfalls of optimizing for engagement, metrics must better account for user intent and well-being while learning their preferences.

\para{Are user preferences shaped by design?}
76\% of video views on YouTube are driven by recommendations. Revealed preferences are learned when a user makes a selection among a set of options. Through recording the users engagement history, the algorithm learns their preferences. Therefore, taken together—the recommendations themselves indirectly influence the revealed preferences of users as they shape what choices users are given, creating \textit{constructed preferences} \cite{thorburn_what_2022}. Furthermore, the very design and affordances of the platform itself exert a significant, often overlooked, influence on user behavior. Prior research \cite{massanari_gamergate_2017}, discusses how platform affordances, such as those on Reddit, shape user interactions and consumption patterns. Building on this, Alfano \etal \cite{alfano_technologically_2021} introduce the concept of "technological seduction," demonstrating how platforms like Google and YouTube strategically cultivate user trust through design and algorithmic curation, leading users to believe the platform understands their desires better than they do themselves. In doing so users are driven into curated bubbles that reinforce users biases, nudged by the algorithm. These ideas corroborate with studies that indicate how altering rankings or recommendations can steer user consumption behaviors \cite{yu_nudging_2024}. Given this ability of platforms to nudge user preferences, the widespread affordances and design of modern platforms indicate that platforms often nudge users to make quick, low-effort decisions. Such decisions reveal impulsive or short-term preferences. Autoplay, Up Next panels \cite{lukoff_how_2021}, infinite scroll, and short video feeds are all examples of design features that capitalize on minimal cognitive friction. These affordances are observed in tandem with phenomena such as `doomscrolling' \cite{baughan_i_2022} or consumption of so-called `brainrot' content \cite{news_what_2024}—content that is considered low quality engaging content but provides no value. While it is technically true the algorithm is supplying users what they appear to want, it often does so only after users have been nudged into a cognitively amenable state where they engage more predictably. The popularity of ASMR and other forms of content designed for passive engagement further illustrates this trend.


\subsection{Limitations}
In this section, we highlight the limitations of our work. \label{sec:discussion:limitations}
\subsubsection{Latent treatment effect.}
A key limitation of our design is the potential for a \textit{latent treatment effect}, where the algorithm’s response may reflect correlations with sub-topics, creators, or niches rather than the sock puppet’s assigned emotional preference. This complicates the interpretation of whether the algorithm is truly learning and reinforcing the emotional preference or simply responding to a correlated characteristic. To mitigate this, we employ multiple seed videos—including both topical and randomly sampled content—to ensure that observed effects are not confined to specific sub-topics. Stable patterns across diverse seeds and iterations strengthen the validity of our conclusions, though complete disentanglement of these factors remains a challenge.

\subsubsection{Sock puppet methodology.}
The sock-puppet methodology, while a widely accepted audit approach, inherently simplifies the behavior of real users. Real users have more complex preferences and are influenced by a number of factors which are not accounted for in this study. However, we posit that demonstrating an impulse response through controlled behavioral signals is valuable in understanding how the algorithm operates—regardless of the validity of the inputs—allowing us to commenting on the behavior of the algorithm rather than its responses.

\subsubsection{Limited interaction signals.}
The absence of other behavioral signals (e.g., likes, dislikes, comments, and search behaviors) in the sock-puppet accounts limits the richness of the data that might otherwise influence algorithmic recommendations. We are limited to basic affordances (such as adding a video to watch history) due to the limitations of the API used. The omitted behaviors likely play a significant role in shaping the recommendations real users receive, potentially leading to differences in the reinforcement patterns observed in this study.

\subsubsection{Dependence on text analysis in utility functions.}
The utility functions used in this study focus exclusively on the textual content of video transcripts to evaluate alignment with emotional preferences. While linguistic cues are a significant component of emotional content, they do not capture the full range of multimodal signals, such as tone, visual imagery, background music, or even non-verbal cues like facial expressions. Additionally, tools used to measure emotion and sentiment can sometimes misclassify subtle or context-dependent language, leading to potential inaccuracies in measuring utility. However, these simple tools provide a scalable and quantifiable measure of emotional alignment. Since both treatment and control groups are subject to reliance on textual analysis, their differences may effectively cancel out any biases introduced by this approach.
\balance


\bibliographystyle{ACM-Reference-Format}
\bibliography{youtube}

\appendix
\section{Appendix}
\subsection{Validating the Innertube API}
\label{sec:appendix:validation}
Prior works auditing YouTube's algorithm using sock puppets have been limited by their ability to create accounts and perform actions repeatedly. These methodologies rely on instrumenting web browsers, most commonly through tools such as selenium, to interact with the platform: communicating inputs and observing the responses. However, due to bot detection techniques employed by platforms, performing large scale audit studies is monetarily expensive and time-consuming due to the resources required to circumvent these measures.
Innertube is YouTube's internal, unified API that offers access to core functionalities, including the ``Up Next'' recommendations \cite{stone_how_2015}. We used this API to collect recommendations for our experiments and sought to validate that these API-based, personalized recommendations are consistent with those served via YouTube’s desktop interface.

To conduct this validation, we created six user accounts, each reflecting one of three preferences: \emph{conservative}, \emph{liberal}, and \emph{conspiracist}. Each preference was trained on 25 videos obtained from TheirTube,\footnote{\url{https://www.theirtube.co/}} a project that curates video lists reflecting specific personas. With two sock puppet accounts per preference, one was trained via the Innertube API and the other via the desktop interface. After training, each sock puppet was used to collect personalized recommendations for 5 new videos while watch history was disabled on the desktop side. We denote a sock puppet with preference \(p\) (e.g., conservative, liberal, conspiracist) and condition \(c\) (e.g., trained via API or desktop) as \(S_p^c\). For each sock puppet, we gathered its recommendation sets \(R_i^c\) from both the Innertube API and the desktop interface, repeating the process three times for each.

We then measured recommendation similarity using the sets of recommended video identifiers. Specifically, we compared:
\begin{itemize}
    \item Within-group similarities: between sock puppets sharing the same condition and preference,
    \item Across-condition similarities: between sock puppets sharing the same preference but trained or observed on different conditions,
    \item Across-preference similarities: between sock puppets trained on different preferences.
\end{itemize}

We found that within-group similarities exceeded across-group (preference) similarities, while remaining roughly equivalent when we only changed the training or observation condition. These results indicate that recommendations from the Innertube API are comparable to those obtained via YouTube's desktop interface for the same user preferences.

Table~\ref{tab:dummy_similarities} provides a simplified view of these comparisons. The similarity values illustrate that the Innertube API serves recommendations that closely match those shown by the desktop app when controlling for user preference.

\begin{table}[H]
\caption{Table illustrating mean recommendation similarity (higher is more similar).}
\label{tab:dummy_similarities}
\begin{tabular}{lcc}
\toprule
\textbf{Groups Compared} & \textbf{Mean Similarity} & \textbf{Std. Dev.} \\
\midrule
Within-group  & 0.43 & 0.06 \\
Across-preferences  & 0.11 & 0.05 \\
Across-conditions  & 0.41 & 0.08 \\
\bottomrule
\end{tabular}
\end{table}

\begin{table}[t!]
\small
\caption{A sample of seed videos from each category.}
\centering
\label{tab:videos}
\begin{tabular}{lp{3cm}p{2cm}}
\toprule
\textbf{Video ID} & \textbf{Title} & \textbf{Channel} \\
\midrule
\multicolumn{3}{l}{\textbf{Fitness}} \\
\texttt{iHrWS86nyXc} & The Only 3 Chest Exercises You NEED to Build a FULLER Chest & Tyler Path \\
\texttt{KIl70ffF5FM} & The ONLY 3 Chest Exercises You Need To Build Muscle (Dumbbells Only!) & Gravity Transformation\\
\texttt{XoTXuZQ4SaU} & CBUM: Chest Workout for Mass (Full Workout) & Chris Bumstead \\
\texttt{NsEbXsTwas8} & The Best and Worst Chest Exercises To Build Muscle (Ranked!) & Jeremy Ethier \\
\midrule
\multicolumn{3}{l}{\textbf{Gaming}} \\
\texttt{E21qSEyRa88} & Gollum is way worse than even our lowest expectations (Review) & IGN \\
\texttt{stjp3h38HCM} & The Top 10 Greatest Plays in Esports History & theScore esports \\
\texttt{11t0xgClpjc} & I Built 4 Secret Rooms You'd Never Find! & Unspeakable \\
\texttt{04E9e1AoJrU} & 10 NEW Game Announcements That Would FREAK Us Out & gameranx \\
\midrule
\multicolumn{3}{l}{\textbf{News}} \\
\texttt{CSwSxWbX5UE} & CNN political commentators clash over Trump's comments & CNN \\
\texttt{ydN79OaYxW0} & CTV National News & CTV News \\
\texttt{bqWwV3xk9Qk} & Why US elections only give you two choices & Vox \\
\texttt{QEmBI5P\_Ld0} & RFK Jr. could win 2024 presidential election: Bo Snerdley & Morning in America \\
\midrule
\multicolumn{3}{l}{\textbf{Random}} \\
\texttt{9xyGp-fNUw4} & Polkadot and The Tale of Blockchain by Irina Karagyaur
 &  Kerala Blockchain Academy \\
\texttt{CEs99-knyFE} & Hot Box At the Shithole 2/20
 & 
Charles Pettitt \\
\texttt{59KVM-jrqMw} & Jackon Review & 
The Black Widow Channel \\
\texttt{8tBkU-guqTE} & A Day in the Life as a D1 Athlete
 & Ben Finneseth \\
\bottomrule
\end{tabular}
\end{table}



\end{document}